\documentclass[prl,twocolumn,
superscriptaddress]{revtex4-1}
\usepackage{amssymb}
\usepackage{amsfonts}
\usepackage{bbm}
\usepackage{mathrsfs}
\usepackage{graphics,graphicx,epsfig,epstopdf,bm,amsmath,amsthm,amssymb}
\usepackage{bm}
\usepackage{bbm}
\usepackage{longtable}
\usepackage{multirow}
\usepackage{array}
\usepackage{color}
\usepackage[usenames,dvipsnames]{xcolor}

\usepackage{float}
\usepackage[a4paper,colorlinks=true,
linkcolor=blue,citecolor=blue,
pdfauthor={ },
pdftitle={ },
pdfsubject={ },
pdfkeywords={ }]{hyperref}

\begin{document}
\title{Observation of Non-Markovianity at Room Temperature by Prolonging Entanglement in Solids}
\author{Shijie Peng}
\author{Xiangkun Xu}
\author{Kebiao Xu}
\author{Pu Huang}
\author{Pengfei Wang}
\author{Xi Kong}
\author{Xing Rong}
\author{Fazhan Shi}

\affiliation{CAS Key Laboratory of Microscale Magnetic Resonance and Department of Modern Physics, University of Science and Technology of China (USTC), Hefei, 230026, China}
\affiliation{Hefei National Laboratory for Physics Sciences at
Microscale and Department of Modern Physics, USTC, Hefei, 230026, China}
\affiliation{Synergetic Innovation Center of Quantum Information and Quantum Physics, USTC, Hefei, 230026, China}
\author{Changkui Duan}
\altaffiliation{ckduan@ustc.edu.cn}
\author{Jiangfeng Du}
\altaffiliation{djf@ustc.edu.cn}
\affiliation{CAS Key Laboratory of Microscale Magnetic Resonance and Department of Modern Physics, University of Science and Technology of China (USTC), Hefei, 230026, China}
\affiliation{Hefei National Laboratory for Physics Sciences at
Microscale and Department of Modern Physics, USTC, Hefei, 230026, China}
\affiliation{Synergetic Innovation Center of Quantum Information and Quantum Physics, USTC, Hefei, 230026, China}


\begin{abstract}


The non-Markovia dynamics of quantum evolution plays an important role in open quantum sytem. However, how to quantify  non-Markovian behavior and what can be obtained from non-Markovianity are still open questions, especially in complex solid systems. Here we address the problem of quantifying non-Markovianity with entanglement in a genuine noisy solid state system at room temperature. We observed the non-Markovianity of quantum evolution with entanglement. By prolonging entanglement with dynamical decoupling, we can reveal the non-Markovianity usually concealed in the environment and obtain detailed environment information. This method is expected to be useful in quantum metrology and quantum information science.
\end{abstract}


\maketitle


Open quantum systems are always exposed to an external environment, which result in interacting and exchanging information between quantum systems and their surroundings. The dynamics of real open quantum systems are often expected to deviate from the idealized Markovian process of losing information to their surrounding environment and to exhibit non-Markovian behaviour with information flowing back to the quantum system from the environment \cite{bookopensystem}. Non-Markovian dynamics is responsible for a wide variety of interesting systems including quantum optics \cite{bookopensystem}, solid systems \cite{quantumdot,semiconductor,solid}, and even some problems in quantum chemistry \cite{pleniormp} and biology systems \cite{biology}. In recent years, more and more attentions have been paid to non-Markovian processes in theory \cite{cirac,piilo,plenioprl,luoshunlong,nori,prldegree,review,plenioassistedentanglement,pleniometrology,vasile}, with some experimental characterization on quantum optics \cite{lichuanfengnaturephysics,chiuri,fanchini}. Due to the ability of regaining lost information and recovering coherence, the Markovian dynamics also show important application prospects in quantum metrology \cite{pleniometrology} and quantum key distribution \cite{vasile}.


In order to clearly distinguish the regimes of Markovian and non-Markovian quantum evolutions and to quantify memory effects in the open system dynamics, the measure for the degree of quantum non-Markovianity has been introduced \cite{cirac}. Several methods based on semigroup, divisibility or flow back of information, and quantum mutual information for the measurement of non-Markovianity have been developed recently \cite{piilo,plenioprl,luoshunlong,prldegree}. And some experiments have been done in quantum optics systems  according to these methods through changing the mimic external environment with the knowledge of model of environment \cite{lichuanfengnaturephysics,chiuri,fanchini}. Under the situation of the absent of an accurate microscopic model of the system-bath interaction, which may actually be unfeasible especially in many body systems, and in order to avoid the definition of an optimization problem \cite{piilo}, entanglement was introduced to measure deviations from Markovianity \cite{plenioprl}. However, in real quantum systems, especially in solids, entanglement are fragile and influenced by decoherence due to the inhomogeneous noise in the surrounding environment, so the non-Markovianity is usually concealed in the Markovian behavior and has not been observed yet with entanglement. Nowadays dynamical decoupling \cite{viola,nature2009} has been used to suppress the inhomogeneous noise and prolong the entanglement coherence time in realistic solid systems \cite{wangya}, from which we could observe and study the degree of non-Markovianity of the quantum system evolution.

In recent years, Nitrogen Vacancy (NV) center system in diamond has attracted more and more attentions due to its promising potential in quantum metrology \cite{magnetic1,electronic,thermal} and quantum information processing \cite{fazhan,Hansonphotonentanglement} at room temperature.
As the open quantum system in solids, it is interesting itself to gain a clear knowledge of its quantum dynamical evolution and it is also very important to know the quantum dynamical properties of NV center quantum system for its application in quantum control and quantum metrology. Here in this letter, we present an experimental study of non-Markovianity in diamond solid system. With the entanglement of single electron spin of NV center in diamond and its nearby ancillary nuclear spin, we have seen the concurrence revival of these two qubits entanglement which reveals the non-Markovianity of the NV center quantum evolution. By applying dynamical decoupling pulses on the single electron spin, we observed the non-Markovianity and obtained more detailed information of the environment memory effect influencing the quantum non-Markovianity.

In the experiment, a single NV center coupled to a first shell $^{13}\rm C$ nuclei in diamond is chosen to study the non-Markovianity of quantum evolution. The diagram of the system is shown in fig.\ 1(a). The direction of the external magnetic field to adjusted to be along the symmetry axis, and the Hamiltonian of the NV center electron spin with a $^{13}\rm C$ nuclear spin can be described as:
\begin{equation}
\label{eq£º1}
 H = \gamma_eB_zS_z+DS_{z}^{2}+\hat{S}\tilde{A}\hat{I}+\gamma_cB_zI_z.
\end{equation}
The zero-field splitting with $D=2.87$ GHz, the Zeeman term with $\gamma_e=2.802$ MHz/Gauss of the electron spin and $\gamma_c=1.071$ KHz/Gauss of the $^{13}\rm C$ nuclear spin and the hyperfine coupling tensor $\tilde{A}$ determine the energy level structure shown in Fig.\ 1(b). The detailed hyperfine coupling tensor $\tilde{A}$ can be found in reference \cite{coupling}. The hyperfine coupling term between the NV center electron spin and the intrinsic Nitrogen nuclear spin is not shown here. During the experimental process, we mainly consider the entanglement of the electron spin and nuclear spin in the subspace of $M_s=0$ and $M_s=1$ of the electron spin which is shown in Fig.\ 1(b) by orange lines. In type IIa diamond, the environment of the NV center is mainly the surrounding $^{13}\rm C$ nuclear spin bath, which brings about the loss of NV center's quantum coherence \cite{huangpu}. As the direct dipole-dipole coupling between two nuclear spins is weak, a $^{13}\rm C$ nuclear spin on the first shell of the NV center can be adopted as an ancilla qubit for the NV electron spin to form the two-qubit system of interest depicted in Fig.\ 1(b), with state $|{M_S=0,M_I = 1}\rangle$ denoted as $|01\rangle$ and alike.


\begin{figure}[htpb]\centering
\includegraphics[width=1\columnwidth]{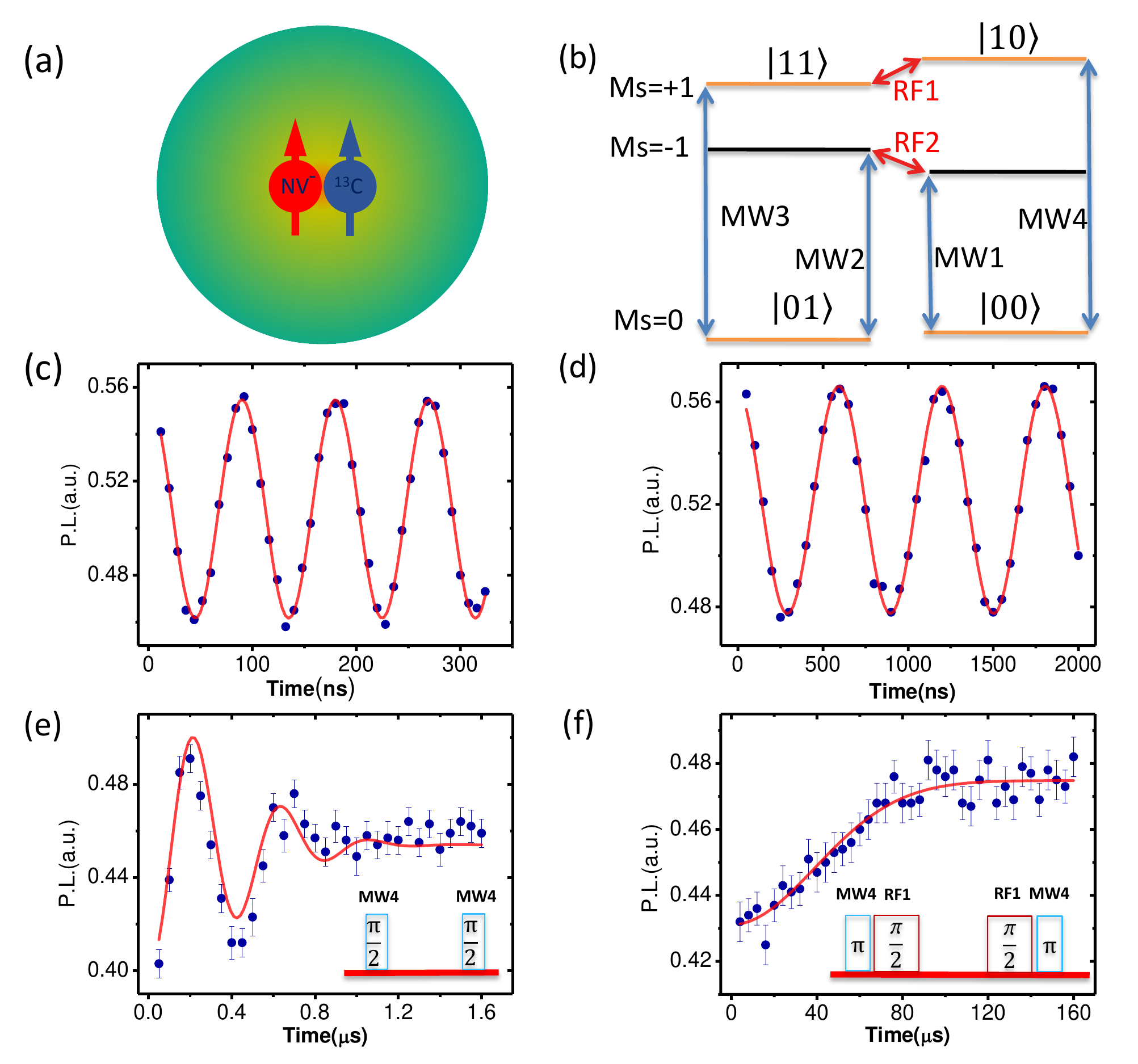}
  \caption{Experimental system.
      (a) schematic of the studied quantum system composing of a NV center and nearby $^{13}\rm C$ nuclei in spin bath.
      (b) energy level scheme of the system. The four states used here are from the subspace containing $M_s=0$ and $M_s=1$ states (labeled $|00\rangle$, $|01\rangle$, $|10\rangle$ and $|11\rangle$). MW1 to MW4 stand for the the transitions between electron spins, and RF1 and RF2 represent the transitions of nuclear spin.
       (c) and (d) are the electron and nuclear spin Rabi nutations, respectively. (e) and (f) are the free induction decays of electron and nuclear spins, respectively.
}
  \label{fig1}
\end{figure}

All measurements were carried out under ambient conditions on a type IIa bulk diamond sample, in which $^{13}\rm C$ has the natural abundance(1.1$\%$) and the nitrogen impurity concentration is less than 5 ppb. Single NV center was addressed by a home-built confocal microscope. The microwave fields used to control the electron spin were generated from Apsin 6000 Signal Generators passed through a switch (Mini-Circuits ZASWA-2-50R+) and a 16W power amplifier. The radio-frequency fields for the excitation of nuclear spin were generated by a direct digital synthesis and went through a switch and a phase shifter (carefully tuned at $90^{o}$), and amplified (Mini-Circuits ZHL-20W-13+) before combining with microwave signals. The combined signals through a diplexer were transmitted to the the NV center through a coplanar waveguide beneath the diamond. Three pairs of adjustable Helmholtz coils were used to generate a 60 Gauss magnetic field to remove the degeneracy of $M_{s}=\pm 1$ state. In the experiment, the length of initialization laser pulse is $3~\mu$s and the waiting time following the laser is $5~\mu$s. The photoluminescence is measured during an integration time of $0.35\ \mu$s. To suppress the photon statistic error, each measurement is typically repeated more than $10^6$ times.

Fig.~1 shows the basic control abilities and coherent properties of the system. The resonance frequency of the nuclear spin in the electron spin $M_s=1$ subspace can be obtained from electron nuclear double resonance spectrum. The experimental data is not shown here. From the sequences of free induced decay, which is depicted at the low right corner of Fig.\ 1(e) and Fig.\ 1(f), we find the coherence time of NV center is $T_{2e}^{*}=0.75~\mu$s and the coherence time of the nuclear spin in the electron spin $M_s=1$ subspace is $T_{2n}^{*}=56~\mu$s. The short coherence time of nuclear spin is caused by the interaction with the surrounding nuclear spins mediated by the NV center electron spin.

\begin{figure}[htpb]\centering
\includegraphics[width=1\columnwidth]{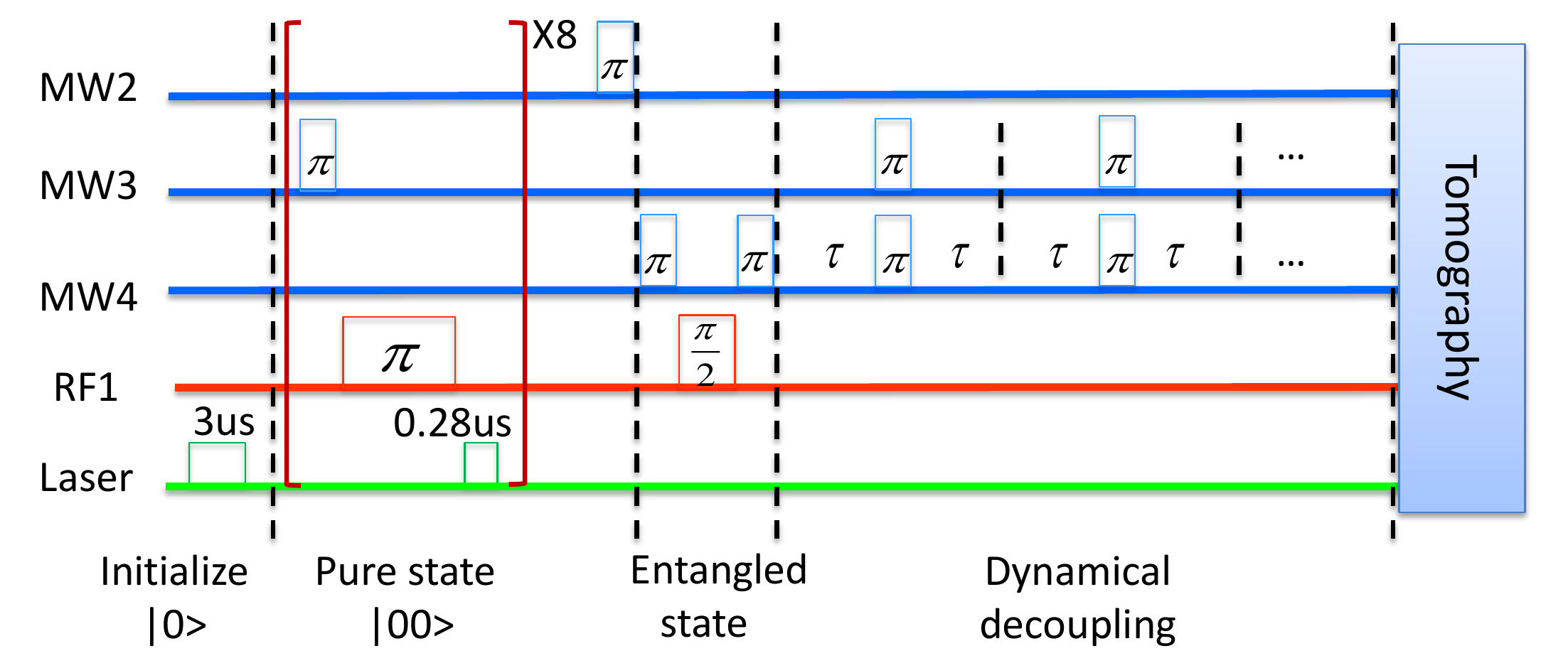}
  \caption{Initialization, dynamical decoupling and measurement of the system.
    To initialize the system into the maximum entangled state ${{1}\over {\sqrt2}}\left (|00\rangle-|11\rangle\right )$, firstly, the electron spin of NV center is transfer to $M_{s}=|0\rangle $ state by a laser pulse lasting $3\ \mu$s, then the system is transfer into $|00\rangle$ state by a pulse sequence of  $8$ times of cycling followed by a MW2 $\pi$ pulse, and finally the combination of two $\pi$ MW4 pulse with one $\pi/2$ RF1 pulse in between completes the preparation of the preparation of the entangled state. Periodic pulsed dynamical decoupling sequence is then exerted to the electron spin through MW3 and MW4 simultaneously. Measurement of the system is done by quantum tomography.
    }
  \label{fig2}
\end{figure}

\begin{figure}[htpb]\centering
\includegraphics[width=1\columnwidth]{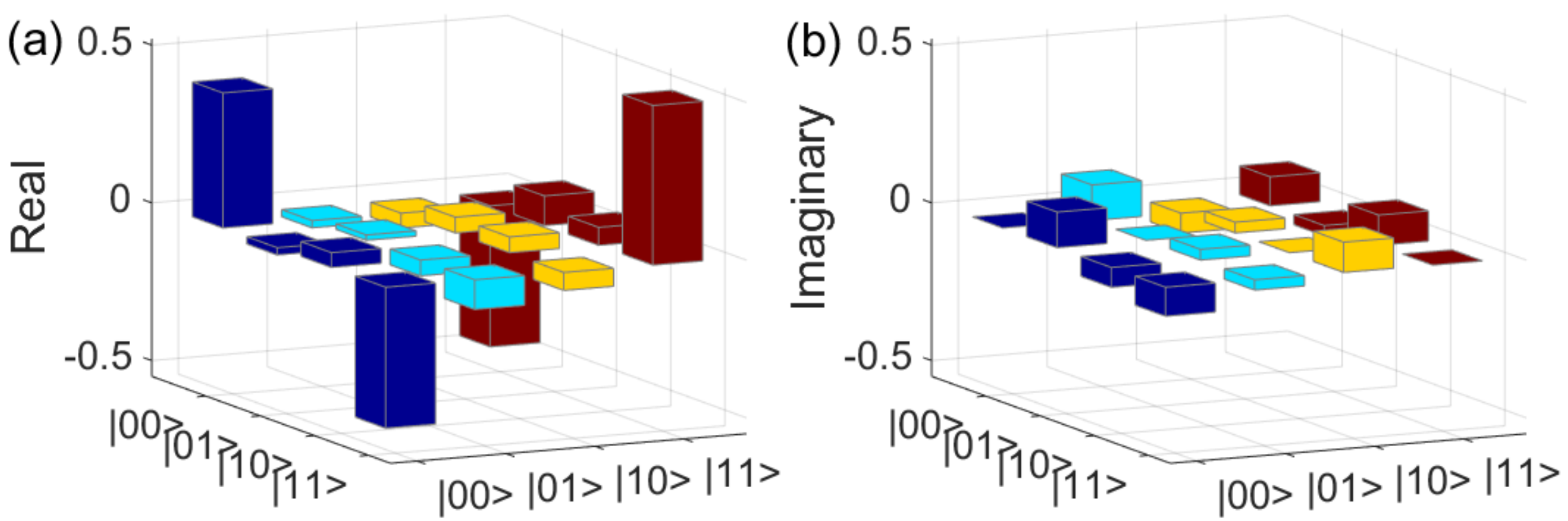}
  \caption{Results of quantum state tomography.
     (a) and (b) are the real and imaginary parts, respectively, of the initial state density matrix. The main off-diagonal elements $|00\rangle\langle11|$ and $|11\rangle\langle00|$ clearly show that the state is $\phi ^{-}=\frac{1}{\sqrt2}\left (|00\rangle-|11\rangle\right )$. The fidelity is obtained as 0.88.}
  \label{fig3}
\end{figure}

The main experimental process is described in Fig.~2. At the beginning, we use cascaded MW, RF and laser pulses to polarize the population to $|00\rangle$ state. This process is mainly through the polarization transfer from the electron spin to the nuclear spin. The sequence is repeated 8 times and the degree of polarization is about 0.8. Then we apply another MW2 $\pi$ pulse to transfer the rest population on $|01\rangle$ state to the $M_s=-1$ subspace, so it is the pure state in the subspace of interest. Then we use the combined Microwave pulses to generate one of the Bell states $\frac{1}{\sqrt2}\left (|00\rangle-|11\rangle\right )$.
Through a period of free evolution or with pulsed dynamical decoupling during evolution, the final state is readout by quantum sate tomography technique. All the density matrix elements are transferred to the $M_{s}=|0 \rangle$ state of the electron spin of the NV center and readout by the fluorescence.

During the experiment of entangle state preparation, the $\pi$/2 pulse was exerted to the nuclear spin to minimize the decoherence influence from the NV center. When doing the quantum state tomography, the transition between $|11\rangle$ and $|10\rangle$ is defined as the working transition, for example, when measuring the density matrix element $|11\rangle\langle10|$, we do the Rabi nutation with two $90^o$ phase shifted radio frequency pulses. Other elements can be measured by transferring the corresponding nutations to the working transition. The population and coherence can be calculated from the rotation curves. Fig.~3 shows the real and imaginary part of the initial state density matrix. The fidelity is estimated by the equation $F = {\rm tr} (\sigma \rho)$, here $\rho$ is our initial state density matrix and $\sigma$ is the ideally expected one, the result we get is 0.88, while the concurrence is about 0.67. Here, the concurrence is calculated by the formula $C(\rho) = \max \{ 0,\lambda_1-\lambda_2-\lambda_3-\lambda_4 \}$ \cite{concurrence}, where the $ \lambda_i$ s are the square roots of the eigenvalues of $ \rho\tilde{\rho} $ in descending order. Here, $\tilde{\rho}=(\sigma_y \otimes \sigma_y)\rho^\ast(\sigma_y \otimes \sigma_y)$, where $\sigma_y$ is the standard Pauli matrix.

\begin{figure}[htpb]\centering
\includegraphics[width=1\columnwidth]{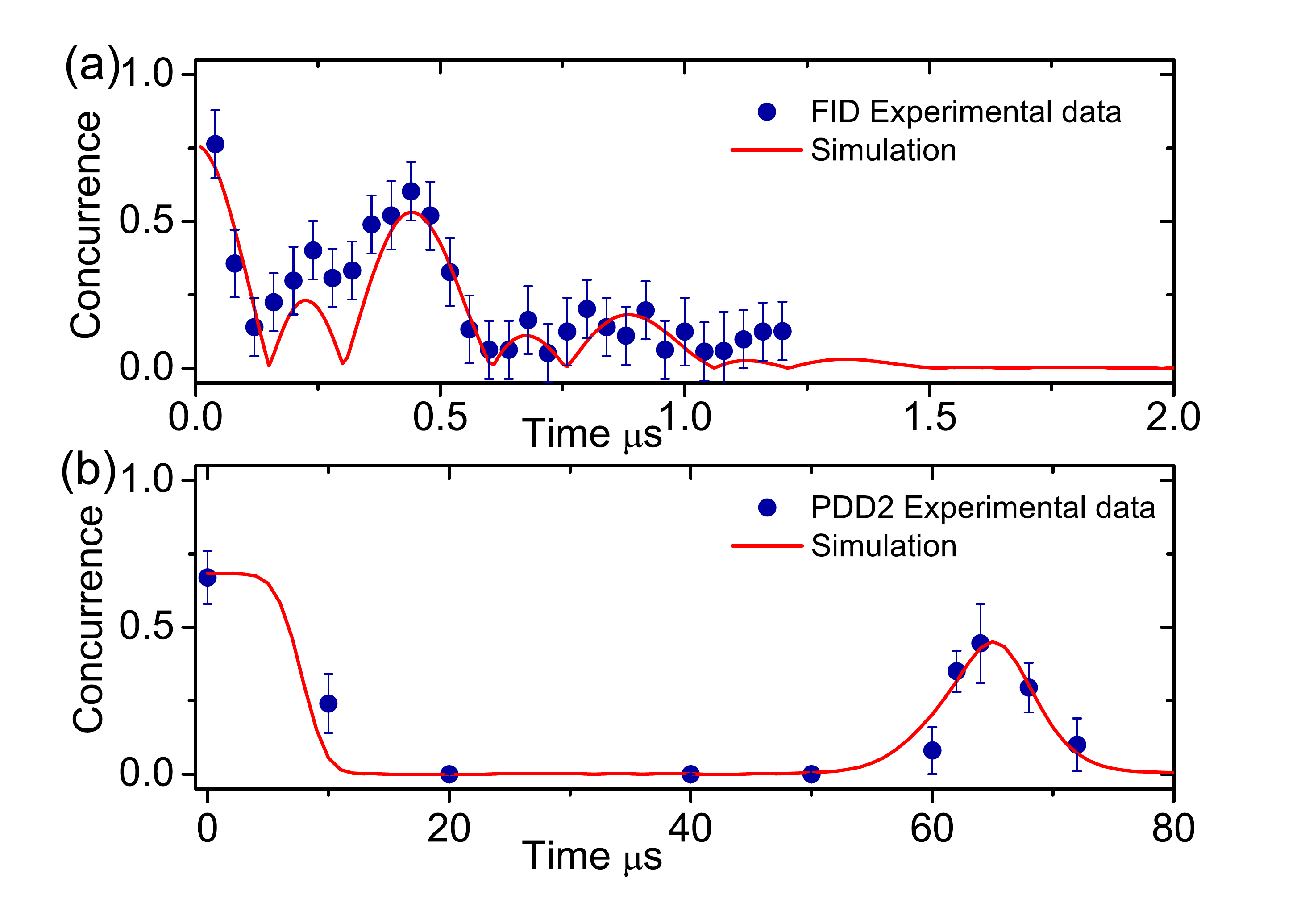}
  \caption{Measured concurrence of entanglement evolution.
 (a) the concurrence of entanglement of free induced decay of the entanglement evolution, which is estimated from the off-diagonal terms of the density matrix.  (b) the concurrence of the entanglement evolution under PDD2 on the electron spin state of NV center, which is calculated from the whole quantum state tomography. The blue dots are experimental data, with the error bars being estimated via Monte Carlo method. The red line is simulation by utilizing cluster correlation expansion \cite{cce} for the sample with natural abundance of $^{13}{\rm C}$. }
  \label{fig4}
\end{figure}

In Fig.~4, we study the non-Markovianity of the quantum evolution with entanglement. From the simple way to quantify the degree of non-Markovianity of quantum evolution introduced in paper\cite{plenioprl},the non-Markovianity within a selected interval $[t_0, t_{\rm max}]$ can be given by:
\begin{equation}
\label{eq£º2}
 \emph{I}^{(E)}= \int_{t_0}^{t_{\rm max}} \left| \frac{dE[\rho(t)]}{dt} \right| dt - \Delta E.
\end{equation}
Here, $\Delta E=E[\rho(t_0)-\rho(t_{\rm max})]$ and E denotes some entanglement measure. In this paper, E means the entanglement concurrence. From the simple formula, we can see that the non-Markovianity could be gained through the height of the entanglement concurrence revival. We first observed the free evolution of the entangled state in Fig.~4(a), from which we could clearly see the concurrence oscillation during the evolution. This is attributed to the strong coupling between the NV center and N nuclear spin which obviously deviate the Markovian process assumption. The decay is mainly caused by the thermal noise of the surrounding $^{13}\rm C$ nuclear spin bath. In Fig.~4(a), the initial concurrence less than a perfect Bell state is mainly caused by the imperfection of the entanglement preparation pulses. We can see the simulation results match the experimental results by considering these errors. From the non-Markovianity of the free entanglement evolution we could get the coupling strength of N nuclear spin with the quantum system and the thermal noise strength of the nuclear spin bath. 
By using dynamical decoupling methods, this noise could be suppressed and the entanglement would be prolonged. Fig.~4(b) shows the experimental results of entanglement evolution under PDD2 sequences working only on the NV's electronic spin qubit through MW3 and MW4 simultaneously. The entanglement is obviously prolonged and the concurrence of entanglement firstly decays to zero, remains zero and then suddenly revives. The horizon time axis is the total entanglement evolution time. This revival is caused by the environmental nuclear spins collective evolution due to the very weak coupling between nuclear spins in the environment, the nuclear spin bath evolution is mainly influenced by the NV center state and the external static magnetic field. By using the PDD2 pulse on NV center, the coupling of the environmental nuclear spin with the NV center is suppressed. Then the nuclear spin bath evolution is mainly subject to the external magnetic field, which induces the collective evolution of the nuclear spin bath. So the entanglement will be recovered when the nuclear spins simultaneously evolve to their initial states. However, the concurrence of the entanglement did not reach its initial value. One reason is the remaining decoherence of the electron spin under PDD2 caused by the nuclear spin-spin interaction within the environment, which can lead to incomplete recovery of the environmental nuclear spins to their collective initial state; and another reason is due to the limited ancillary nuclear spin coherence time. Here, it is mainly caused by the latter, as usually the nuclear spin-spin interaction is too weak to take effect in the considering time scale\cite{childress}. The ancillary nuclear spin interacts with the environment mediated by the NV center that induced the main decay. By excluding this reason, the non-Markovianity in Fig.~4(b) mainly reflect the interaction strength inside the nuclear spin bath which concern with the bath memory time. Comparisons of Fig.~4(a) with Fig.~4(b) shows that the more non-Markovianity is revealed by dynamical decoupling, which is helpful to understanding the detail origin of non-Markovianity. This method is expected to be used in quantum control and quantum metrology in non-Markovian environment\cite{pleniometrology}.

In summary, we have studied the non-Markovian characteristics of quantum evolution in diamond experimentally by using the entanglement method. We find different non-Markovianity of the entanglement evolutions and provide corresponding explanations. By using the dynamical decoupling, more non-Markovianity is revealed that otherwise is hiding in the environment. From such an experimental non-Markovianity study, we can obtain much more detail interaction information of the environment with the NV center in the diamond. This will be helpful for making non-Markovianity as a resource for quantum technology applications and also would be expected useful in other open quantum systems like transport processes in biological aggregates and complex nano-structures\cite{exiton}.

Acknowledgement--The authors thank Martin B. Plenio for helpful discussions. This work was supported by the National Natural Science Foundation of China (Grants No. ~81788101, No. ~11227901, No. ~11722544, No. ~91636217), the CAS (Grants No. ~ GJJSTD20170001 and No. ~ QYZDY-SSW-SLH004),  the National Key R\&D Program of China (Grant No.~ 2013CB921800, No.~ 2016YFA0502400).




\begin{thebibliography}{99}
\bibitem{bookopensystem}H. P. Breuer and F. Petruccione, {\it The Theory of Open Quantum Systems} (Oxford University Press, 2007).
\bibitem{quantumdot} Y. Kubota and K. Nobusada, J. Phys. Soc. Jpn. \textbf{78}, 114603 (2009).
\bibitem{semiconductor} B. E. Kane, Nature \textbf{393}, 133 (1998).
\bibitem{solid} C. W. Lai, P. Maletinsky, A. Badolato, and A. Imamoglu, Phys. Rev. Lett. \textbf{96}, 167403 (2006).
\bibitem{pleniormp} M. B. Plenio and P. L. Knight, Rev. Mod. Phys. \textbf{70}, 101 (1998).
\bibitem{biology} P. Rebentrost and Aspuru-Guzik, J. Chem. Phys. \textbf{134}, 101103 (2011).
\bibitem{review}{\'A}. Rivas, S. F. Huelga, and M. B. Plenio, Rep. Prog. Phys. \textbf{77}, 094001 (2014).
\bibitem{cirac} M. M. Wolf, J. Eisert, T. S. Cubitt, and J. I. Cirac, Phys. Rev. Lett. \textbf{101}, 150402 (2008).
\bibitem{piilo} H.-P. Breuer, E.-M. Laine, and J. Piilo, Phys. Rev. Lett. \textbf{103}, 210401 (2009).
\bibitem{plenioprl} {\'A}. Rivas, S. F. Huelga, and M. B. Plenio, Phys. Rev. Lett. \textbf{105}, 050403 (2010).
\bibitem{luoshunlong} Shunlong Luo, Shuangshuang Fu, and Hongting Song, Phys. Rev. A \textbf{86}, 044101 (2012).
\bibitem{nori}W.-M. Zhang, P.-Y. Lo, H.-N. Xiong, M. Wei-Yuan Tu, and F. Nori, Phys. Rev. Lett. \textbf{109}, 170402 (2012).
\bibitem{prldegree}D. Chru{\'s}ci{\'n}ski and S. Maniscalco, Phys. Rev. Lett. \textbf{112}, 120404 (2014).
\bibitem{plenioassistedentanglement} S. F. Huelga, {\'A} . Rivas, and M. B. Plenio, Phys. Rev. Lett. \textbf{108}, 160402 (2012).
\bibitem{pleniometrology} A. W. Chin, S. F. Huelga, and M. B. Plenio, Phys. Rev. Lett. \textbf{109}, 233601 (2012).
\bibitem{vasile} R. Vasile, S. Olivares, M. G. A. Paris, and S. Maniscalco, Phys.Rev. A \textbf{83}, 042321 (2011).
\bibitem{lichuanfengnaturephysics} B.-H. Liu, L. Li, Y.-F. Huang, C.-F. Li, G.-C. Guo, E.-M. Laine, H.-P. Breuer, and J. Piilo, Nature Phys. \textbf{7}, 931 (2011).
\bibitem{chiuri} A. Chiuri, C. Greganti, L. Mazzola, M. Paternostro, and P. Mataloni, Scientific Reports \textbf{2}, 968 (2012).
\bibitem{fanchini}F.F. Fanchini, G. Karpat, B. Cakmak, L.K. Castelano, G.H. Aguilar, O. Jim{\'e}nez Far{\'i}as, S.P. Walborn, P.H. Souto Ribeiro, and M.C. de Oliveira, Phys. Rev. Lett. \textbf{112}, 210402 (2014).
\bibitem{viola} L. Viola, E. Knill, and S. Lloyd, Phys. Rev. Lett. \textbf{82} 2417 (1999).
\bibitem{nature2009} J. Du, X. Rong, N. Zhao, Y. Wang, J. Yang and R. B. Liu, Nature \textbf{461}, 1265 (2009).
\bibitem{wangya} Y. Wang, X. Rong, P. Feng, W. Xu, B. Chong, J. Su, J. Gong, and J. Du, Phys. Rev. Lett. \textbf{106}, 040501 (2011).
\bibitem{magnetic1} J.R. Maze, P.L. Stanwix, J.S. Hodges, S. Hong, J.M. Taylor, P. Cappellaro, L. Jiang, M.V.G. Dutt, E. Togan, A.S. Zibrov, A. Yacoby, R.L. Walsworth, and M.D. Lukin, Nature, \textbf{455}, 644 (2008).
\bibitem{electronic} F. Dolde, H. Fedder, M. W. Doherty, T. Noobauer, F. Rempp, G. Balasubramanian, T. Wolf, F. Reinhard, L. C. L. Hollenberg, F. Jelezko and J. Wrachtrup, Nature Phys. \textbf{7}, 459 (2011).
\bibitem{thermal} D. M. Toylia, C. F. de las Casasa, D. J. Christlea, V. V. Dobrovitskib, and D. D. Awschaloma, Proc. Natl. Acad. Sci. \textbf{110}, 8417 (2013).
\bibitem{fazhan} F. Shi, X. Rong, N. Xu, Y. Wang, J. Wu, B. Chong, X. Peng, J. Kniepert, R. Schoenfeld, W. Harneit, M. Feng, and J. Du, Phys. Rev. Lett. \textbf{105} 040504 (2010).
\bibitem{Hansonphotonentanglement} H. Bernien, B. Hensen, W. Pfaff, G. Koolstra, M. S. Blok, L. Robledo, T. H. Taminiau, M. Markham, D. J. Twitchen, L. Childress and R. Hanson, Nature \textbf{497}, 86 (2013).
\bibitem{coupling}S. Felton, A. M. Edmonds, M. E. Newton, P. M. Martineau, D. Fisher, D. J. Twitchen, and J. M. Baker, Phys. Rev. B \textbf{79}, 075203 (2009)
\bibitem{huangpu} P. Huang, X. Kong, N. Zhao, F. Shi, P. Wang, X. Rong, R.B. Liu and J. Du, Nature Communications \textbf{2}, 570 (2011).
\bibitem{concurrence} W.K. Wootters, Phys. Rev. Lett. \textbf{80}, 2245 (1998)
\bibitem{childress} L. Childress, M. V. Gurudev Dutt, J. M. Taylor, A. S. Zibrov, F. Jelezko, J. Wrachtrup, P. R. Hemmer, and M. D. Lukin, Science \textbf{314}, 5797 (2006).
\bibitem{exiton}E.Collini and G. D. Scholes, Science \textbf{323}, 369 (2009)
\bibitem{cce}W. Yang and R. B. Liu, Phy. Rev. B \textbf{b}, 085315 (2008)
\end{thebibliography}
\end{document}